\documentclass[11pt]{article}
\usepackage{hyperref}
\pdfoutput=1
\begin{document}
\title{Hovering UFO Droplets}
\author{Sushant Anand, Adam Paxson, Rajeev Dhiman, J. David Smith, 
\\ and Kripa K. Varanasi \\
Massachusetts Institute of Technology, Cambridge, MA 02139, USA} \maketitle
\begin{abstract}
This fluid dynamics video is an entry for the Gallery of Fluid Motion of the 65th Annual Meeting of the APS-DFD. This video shows behavior of condensing droplets on a lubricant impregnated surface and a comparison with a superhydrophobic surface. On impregnated surfaces, drops appear like UFOs hovering over a surface. The videos were recorded in an Environmental SEM and a specially built condensation rig.
\end{abstract}

\section{Introduction}

Superhydrophobic surfaces show excellent non-wetting properties. However during condensation, condensing drops grow in Wenzel pinned state. By impregnating the solid surface with a lubricant this limitation can be overcome. Fluid dynamics videos show condensation of water on conventional superhydrophobic surface and lubricant impregnated surface.

\section{Experiment Description}
Condensing surfaces were fabricated using standard photolithography with well-defined cubical microposts of silicon(a= 10$\mu$m,b= 10$\mu$m, and h= 10$\mu$m) and these were etched further to produce nanograss features. Subsequently they were silanized using a low-energy silane to render them hydrophobic. The textured surface was impregnated with an ionic liquid ([BMIm$^+$][Tf$_2$N$^-$]) by using a controlled dip coating procedure. Experiments for observing microscale condensation growth were performed in an Environmental SEM. The surfaces were titled by 15$^\circ$ to the vertical and the ESEM experiments were conducted under identical conditions (pressure 1000 Pa, substrate temperature ∼4.5 $^\circ$C, beam voltage 25 kV, and beam current 1.7 nA).\ 
Subsequently macroscopic shedding behavior was studied on the same samples in a custom built condensation rig. These experiments were performed with saturated steam (60 kPa, 86 $^\circ$C) at a constant surface cooling flux (160 kW/m$^2$). The condensing droplets were photographed using a digital video camera (Nikon D300S, 24 fps) equipped with a macro lens system (Nikon 105 mm with two 2x teleconverters).

\section{Description}
On superhydrophobic surfaces, droplets rest atop surface textures in a Cassie state and can be shed easily due to reduced pinning. However, as shown in the video, during condensation, the droplets nucleate within the surface and subsequently grow and displace the entrained air to remain in an impaled Wenzel state.
This wenzel pinning can however be eliminated by introduction of a lubricant.\cite{quere2005, aizenslips2011, quereslip2011, jdsmith2011} We show that drops condense on impregnated surfaces in Cassie state and microscale drops even with low contact angle move across the surface giving an appearance of hovering UFOs. The lubricant forms a wetting ridge around a water droplet due to interfacial force balance at the three phase contact line. The shedding behavior of droplets on such surface is studied macroscopically and compared with conventional superhydrophobic surfaces. The remarkable shedding behavior seen on these surfaces make them promising for enhanced condensation heat transfer.\cite{sushantnano2012}

Two sample videos are provided (a High Resolution MPG as Video 1 and a Low Resolution MPG as Video 2).

\end{document}